# A Text Analysis of Federal Reserve meeting minutes

**Harish Gandhi Ramachandran**
Department of Computer Science
Illinois Institute of Technology
Chicago, IL 60616, USA
hramachandran@hawk.iit.edu

**Dan DeRose Jr**
Stuart School of Business
Illinois Institute of Technology
Chicago, IL 60616, USA
dderose@hawk.iit.edu

## Abstract

Recent developments in monetary policy by the Federal Reserve has created a need for an objective method of communication analysis. Using methods developed for text analysis, we present a novel technique of analysis which creates a semantic space defined by various policymakers' public comments and places the committee consensus in the appropriate location. It's then possible to determine which member of the committee is most closely aligned with the committee consensus over time and create a foundation for further actionable research.

## 1. INTRODUCTION

The Federal Reserve is responsible for setting overnight interest rates in the United States. The level and future direction of these rates are important for pricing all assets and derivatives in the economy [1]. All investors, implicitly or explicitly, are affected by their decisions. Since the Great Financial Crisis, the Federal Reserve has departed from primarily conducting policy by changing the interest rate and has resorted to unconventional monetary policy where communication plays a leading role [2]. This transition can ironically increase the uncertainty about future movements as analysts inherently incorporate their own biases into their analysis [3]. An object method of analysis is required.

Natural language processing and text analysis are tools to create an objective understanding of text. These methods are increasingly being applied to text created by the Fed. The Fed, which meets eight time a year, creates two sets of text data consistently: a statement which is released immediately after a meeting of the Federal Open Market Committee (FOMC) and minutes of those meetings which are released with a three-week lag. Hernandez-Murillo and Shell (2014) [4] investigate the Fed statement to find increasing levels of complexity using the Flesch-Kincaid Grade Level index over time. Acosta and Meade (2015) [5] investigate evolving word usage and the similarity of the statement over time. The problem remains though, what's the best way to predict the future direction of Fed actions? This paper will use a novel text analysis technique to determine which member of the committee has views most like the committee consensus. It's then possible to monitor a few policymakers in between meetings for a potential change in policy. This is an original implementation of the technology not found in the literature and can help investors understand which policymakers to follow closely.

## 2 EXPERIMENTAL AND COMPUTATIONAL DETAILS

### 2.1 Description of Dataset

The Federal Reserve meets eight times a year to determine the direction of monetary policy. This analysis will utilize the minutes from the eight meetings of 2017. Each set of minutes is approximately 8,000-10,000 words and is available in the archives on the Federal Reserve website in PDF format.

The Federal Open Market Committee (FOMC) is the body within the Fed who discusses and decides policy through a vote. The FOMC consists of twelve members: seven from the Board of Governors, the President of the Federal Reserve Bank of New York, and four of the eleven Reserve Bank Presidents who serve one-year terms on a rotating basis [6]. Over the course of 2017, there were several changes to the committee. These policymakers regularly communicate to the public via speeches or interviews. This analysis will use the public comments of each voting member before each Federal Reserve meeting. The table

1 displays policymakers participated in each meeting as well as the date of the speech utilized in this analysis. This data can be found in the archives on the Federal Reserve website as well as in the archives on the various regional district websites in PDF format.

Table 1: Policymaker attendance and date of speech used in analysis

| Meeting Date | Yellen | Dudley | Brainard | Evans | Fischer | Harker | Kaplan | Kashkari | Powell | Quarles | Tarullo |
|---|---|---|---|---|---|---|---|---|---|---|---|
| Jan31/Feb1 | 19-Jan | 1-Jan | 17-Jan | 24-Oct | 21-Nov | 20-Jan | 30-Nov | 18-Jan | 7-Jan | x | 2-Dec |
| Mar 14/15 | 3-Mar | 15-Feb | 1-Mar | 9-Feb | 3-Mar | 28-Feb | 13-Feb | 6-Mar | 22-Feb | x | 2-Dec |
| May2/3 | 3-Mar | 7-Apr | 28-Apr | 29-Mar | 17-Apr | 3-Apr | 13-Feb | 6-Mar | 28-Mar | x | x |
| June 13/14 | 3-Mar | 11-May | 30-May | 25-May | 17-Apr | 23-May | 13-Feb | 6-Mar | 1-June | x | x |
| July 25/26 | 3-Mar | 26-Jul | 13-Jul | 13-Jul | 6-Jul | 11-Jul | 13-Jul | 6-Mar | 6-Jul | x | x |
| Sept 19/20 | 25-Aug | 7-Sep | 5-Sep | 13-Jul | 31-Jul | 8-Sep | 13-Jul | 6-Mar | 30-Aug | x | x |
| Oct31/Nov1 | 15-Oct | 6-Oct | 12-Oct | 25-Sep | x | 17-Oct | 17-Oct | 27-Sep | 12-Oct | x | x |
| Dec12/13 | 15-Oct | 7-Dec | 16-Nov | 15-Nov | x | 1-Dec | 27-Nov | 27-Sep | 2-Nov | 30-Nov | x |

Note: x denotes an inactive participant

## 2.2 Data Preprocessing

Before any analysis can begin, it's necessary to preprocess the data. Initially, each PDF file is converted to a text file. Then, it's required to clean each text file by removing all stop words, stripping whitespace, eliminating common and irrelevant words like attendance lists among other things. It's now possible to put each text file into the relevant document-term vector (DTV) where each value $n_{d,t}$ is the number of times term $t$ is in document $d$.

## 2.3 LSA, Folding-in, and Cosine Similarity

Latent semantic analysis (LSA) is a text analysis technique used to reduce dimensionality. It is possible, for example, to reduce the number of concepts in a document or to find meaning between terms or documents through LSA [7].

The work horse of LSA is singular value decomposition (SVD) [8]. Where $A$ is a DTM matrix, singular value decomposition is written as follows

$$A = U\Sigma V^T$$

where $U$ is a matrix with d rows for the number of documents and k columns for the number of concepts, $\Sigma$ is a list of singular values k which must be inserted into a diagonal matrix, and V is a matrix with t rows for the number of terms and k columns for the number of concepts. Together these objects are equivalent to $A$, the original DTM matrix. Usually, LSA is utilized to reduce the dimensionality of an existing document. This is accomplished through optimizing the size of the diagonal matrix, $\Sigma$ and thus removing some of the singular values. Then, when executing matrix algebra, $A$ will no longer be the original DTM but will reveal meaning between words and topics.

For purposes of this analysis, LSA is the first step in determining who is the most similar voting member of the Federal Reserve to the whole committee. To begin, a speech DTV from each policymaker active at a particular meeting is concatenated into a document-term matrix (DTM). LSA is then utilized to create a semantic space. In this way, each policymakers' views are considered and the semantic space created is as wide and broad as their cumulative perspectives. Traditional SVD as described above is then run on the DTM. This process all occurs before incorporating the relevant set of minutes. This is a necessity as including a set of minutes in this first decomposition will change the shape of the semantic space. This will impact how the semantic space is created and distort our intention of determining who is most similar to the set of minutes. Next, a set of minutes from a Federal Reserve meeting are 'folded in' to this semantic space. This process occurs through a series of transformations.

Mathematically, SVD is first run on the TF-IDF speech DTM where 'S' denotes the speech data:

$$A_S = U_S \Sigma_S V_S^T$$

Next, a minute's document term vector (DTV) is created using only the terms from the speech DTM as mentioned above. Before proceeding, it's necessary to normalize this DTV using the same global weighting scheme as the speech DTM. This means that instead of a standard TF-IDF transformation, the TF component will utilize the minutes DTV while the IDF will utilize the speech DTM [9].

It's now possible to fold the set of minutes into the semantic space created by the speeches. To begin, the minute's vector, $v^T$, is multiplied by $V_S$ and $\Sigma_S^{-1}$ to create a [1xk] vector, $\hat{d}^T$, which relates the minute's document to the concepts k of the semantic space created by the speeches. Then, $\hat{d}^T$ is multiplied by $V_S$ and $\Sigma_S^{-1}$ to create a [tx1] vector, $\hat{m}$ which relates the term count of the minute's document to the term count of the TF-IDF speech DTM. Mathematically,

$$\hat{d} = v^T V_S \Sigma_S^{-1}$$

and

$$\hat{m} = V_S \Sigma_S \widehat{d^T}.$$

After all transformations, $\hat{m}$ is now in the same terms as the speech TF-IDF DTM. It can now be appended to this DTM and analyzed using cosine similarity. Cosine similarity plots each column as a point in a high dimensional space and discovers the angle between that point and the origin [10]. Two points at opposite directions from the origin will have a high degree of separation and less in common or a low correlation. Conversely, two points close together in space will have a low degree of separation or a high correlation.

## 3 RESULTS

Let's review the original DTV of the set of minutes from the December 2017 meeting that is folded into the set of speeches. Next, let's review $\hat{m}$, the DTV after all transformations.

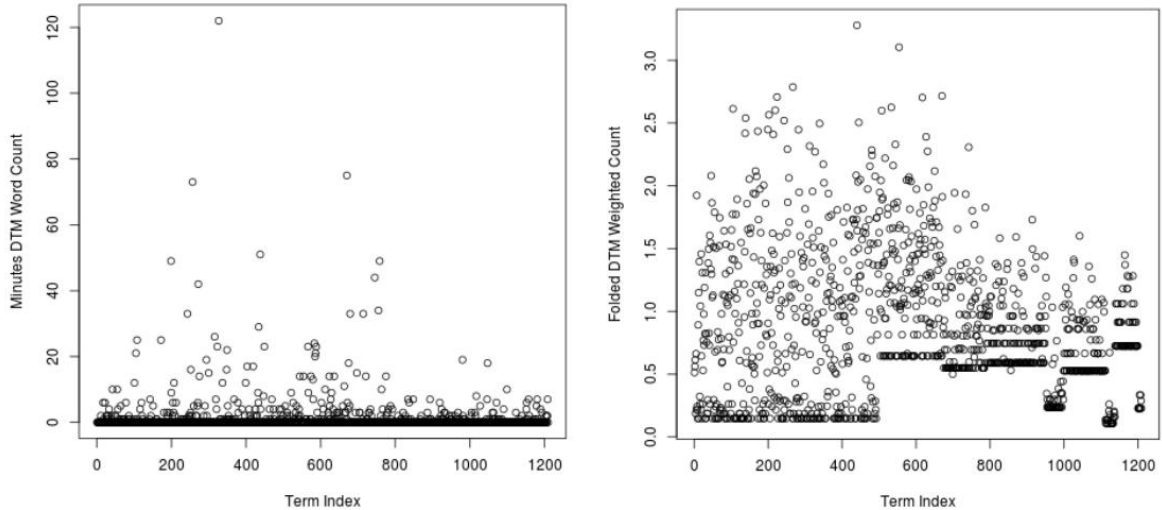

Figure 1 a) Minutes DTM Word Count plot (left) b) Folded DTM Weighted Count (right)

After all transformations, the number of times a word appears is significantly different. The plot shown in Fig. 1a) on the left shows the notional amount of times a word appears; the plot on the right expresses this number in terms of the speech TF-IDF DTM. Each vector can now be appended to its respective speech TF-IDF DTM to discern the correlation between the various policy maker's speeches and the set of minutes. The correlation matrix for the December 2017 meeting is shown in Fig. 2

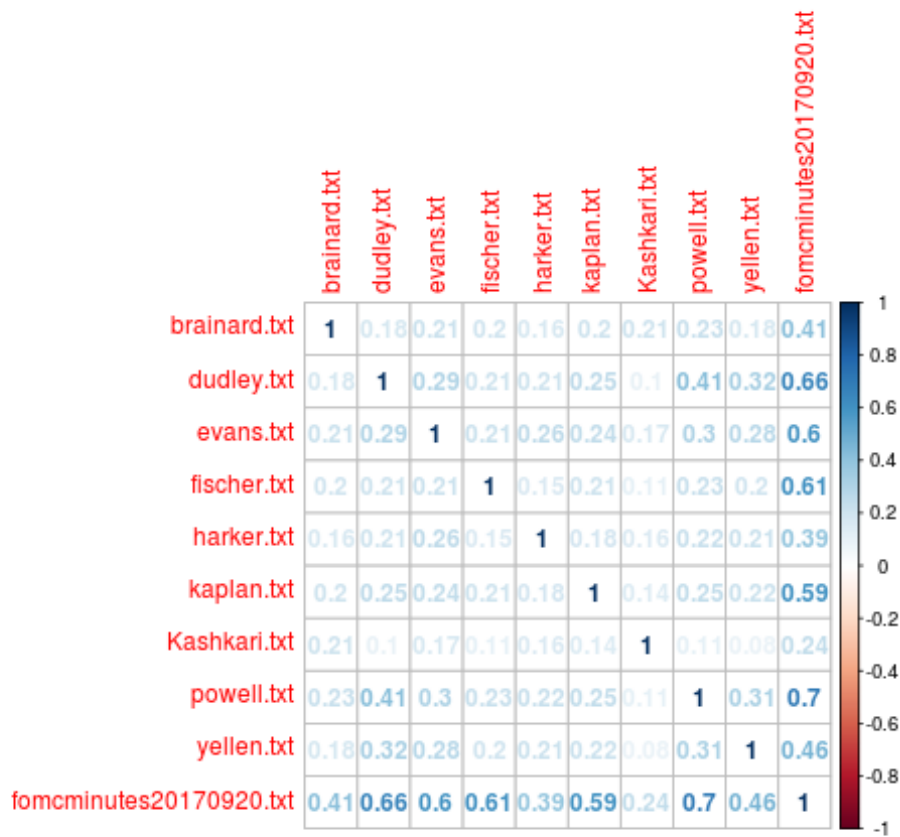

Figure 2 speakers correlation matrix for the December 2017 month.

This process was carried out for each meeting during 2017. The correlation between each policymaker and the minutes is recorded in table 2. It's now possible to discern which policymaker holds views most like the committee consensus. It's also possible to see how those correlations change over time. An active participant in markets can focus their attention on the appropriate policymaker and not be distracted by someone who holds out of consensus views. For example, Neel Kashkari, President of the Minneapolis Fed, regularly holds views that are dissimilar from the committee consensus. Any headlines he creates should not meaningfully impact markets. Conversely, Janet Yellen, who is the Chair of the FOMC, holds a view very similar to the committee consensus. Her views should (and do) have a large impact on markets. Surprisingly, Robert Kaplan, President of the Dallas Fed, holds the most consensus view over the course of 2017. This is a surprising result for market participants. An active participant should carefully watch the evolution of Kaplan's perspective for a leading indicator of a potential change in the FOMC.

This analysis is just an observation of Fed policymaker views in relation to the committee consensus but it can easily be advanced to a deployed state. Further analysis should be directed at the market sensitivity to evolving policymaker views and the permanence of those market movements. A shift in perspective by Kashkari, for example, should not have a lasting impact on markets. A shift by Kaplan or Yellen, however, should have a large and lasting impact. This analysis lays a foundation for further investigation.

Table 2: Correlation between speech and minutes for each meeting

| Meeting Date | Yellen | Dudley | Brainard | Evans | Fischer | Harker | Kaplan | Kashkari | Powell | Quarles | Tarullo |
|---|---|---|---|---|---|---|---|---|---|---|---|
| Jan31/Feb1 | 0.8 | 0.57 | 0.63 | 0.35 | 0.4 | 0.49 | 0.57 | 0.39 | 0.57 | x | 0.57 |
| Mar 14/15 | 0.7 | 0.49 | 0.66 | 0.69 | 0.36 | 0.51 | 0.67 | 0.17 | 0.67 | x | 0.51 |
| May2/3 | 0.74 | 0.52 | 0.39 | 0.73 | 0.58 | 0.47 | 0.66 | 0.17 | 0.38 | x | x |
| June 13/14 | 0.74 | 0.39 | 0.76 | 0.55 | 0.56 | 0.47 | 0.57 | 0.15 | 0.73 | x | x |
| July 25/26 | 0.68 | 0.46 | 0.63 | 0.66 | 0.53 | 0.3 | 0.69 | 0.17 | 0.45 | x | x |
| Sept 19/20 | 0.49 | 0.63 | 0.68 | 0.68 | 0.62 | 0.23 | 0.74 | 0.17 | 0.35 | x | x |
| Oct31/Nov1 | 0.49 | 0.59 | 0.63 | 0.63 | x | 0.37 | 0.7 | 0.35 | 0.62 | x | x |
| Dec12/13 | 0.58 | 0.34 | 0.42 | 0.6 | x | 0.35 | 0.78 | 0.41 | 0.54 | 0.4 | x |
| mean | 0.65 | 0.50 | 0.60 | 0.61 | 0.51 | 0.40 | 0.67 | 0.25 | 0.54 | 0.40 | |
| rank | 2 | 7 | 4 | 3 | 6 | 9 | 1 | 10 | 5 | 8 | |

Note: x denotes an inactive participant

## 3 CONCLUSION

In recent years, the Federal Reserve has transitioned from primarily conducting monetary policy through changing interest rates and incorporated communication policy as a primary tool. This transition requires the introduction of an objective means to analyze those communications. Through LSA, it is possible to create a semantic space of the cumulative perspectives of policymakers of the FOMC. After a series of transformations, it's possible to place a set of minutes into that semantic space. Cosine similarity is a measure of correlation between points in a high dimensional space and can be utilized to determine the similarity between various policymakers and the committee consensus. It's then possible to see objectively which policymaker holds views most like the committee consensus. This analysis can easily be the foundation of a deployed data analytics tool by investigating the sensitivity and permanence of market movements in relation to headlines created by each policymaker.